\documentclass[reprint,twocolumn,groupedaddress]{revtex4-2}
\usepackage[T1]{fontenc}
\usepackage{amsmath}
\usepackage{amssymb}
\usepackage{graphicx}
\usepackage{dcolumn}
\usepackage{booktabs}
\usepackage{bm}

\begin{document}

\title{Recovery of accidental bound states in the continuum in vertically asymmetric photonic lattices through radiation-source parity restoration}

\author{Sun-Goo Lee}
\email{sungooleee@gmail.com}
\affiliation{Department of Data Information and Physics, Kongju National University, Gongju, 32588, South Korea}
\affiliation{Institute of Application and Fusion for Light, Kongju National University, Cheonan, 31080, South Korea}

\author{Wook-Jae Lee}
\email{wookjaelee@kongju.ac.kr}
\affiliation{Department of Data Information and Physics, Kongju National University, Gongju, 32588, South Korea}
\affiliation{Institute of Application and Fusion for Light, Kongju National University, Cheonan, 31080, South Korea}
\date{\today}

\begin{abstract}
Vertical asymmetry generally precludes accidental bound states in the continuum (BICs) in planar photonic lattices, yet the underlying physical mechanism remains unclear. In this study, we show that accidental-BIC formation is governed by the definite parity of the transverse radiation-source profile rather than by geometric vertical mirror symmetry. A semianalytical coupled-mode analysis shows that an accidental BIC occurs when the radiation overlap integral of a leaky Bloch mode vanishes. We decompose the overlap into contributions from the even- and odd-parity components of the transverse radiation-source profile. Complete radiation suppression requires both contributions to vanish simultaneously. In vertically symmetric lattices, the source profile has definite even or odd parity, causing one contribution to vanish identically. Thickness tuning can null the remaining contribution and produce an accidental BIC. In vertically asymmetric lattices, by contrast, both parity components are generally present. The waveguide dispersion forbids their simultaneous vanishing, thereby precluding an accidental BIC. Introducing a parity-restoring layer into a vertically asymmetric lattice restores definite parity to the transverse radiation-source profile, thereby enabling the formation of an accidental BIC. Finite-element simulations confirm the analytically predicted recovery of accidental BICs for both TE$_0$ and TE$_1$ Bloch modes.
\end{abstract}

\maketitle

\section{Introduction}

Optical bound states in the continuum (BICs) are special electromagnetic eigenstates that remain perfectly confined in open photonic structures even though their frequencies lie within the continuous spectrum of radiating waves~\cite{Hsu2016,MKang2023,JYu2025}. BICs are completely decoupled from the surrounding radiation continuum and therefore possess a diverging radiative quality ($Q$) factor and a vanishing resonance linewidth. With highly enhanced radiative $Q$ factors, BICs offer significant advantages for a wide range of applications, including low-threshold lasing~\cite{Kodigala2017,MSHwang2021,SHan2023}, high-sensitivity sensing~\cite{Tittl2018,Yesilkoy2019}, narrowband optical filtering~\cite{Wang1990,MNiraula2015,Doskolovich2019,YLiu2023}, and enhanced nonlinear light–matter interactions~\cite{Koshelev2019,Liu2019,Koshelev2020}.
Different types of BICs have been reported in various photonic systems~\cite{Marinica2008,Bulgakov2008,Plotnik2011,Azzam2018,Minkov2018,Gao2019,SGLee2020Nanoph,SGLee2021PRL,SGLee2021OE,Kang2025Janus,Zuo2025Janus,SGLee2026JOpt}. Among them, accidental BICs in planar photonic structures, including one-dimensional (1D) gratings, two-dimensional (2D) photonic crystal slabs, and metasurfaces, have attracted considerable attention~\cite{Hsu2013,Yang2014,Doeleman2018,Neale2021,Hu2022}. Accidental BICs occur at generic $k$ points and, owing to the conservation of topological charge, are not destroyed by small variations in the structural parameters but instead undergo only slight shifts in momentum space~\cite{BZhen2014,BWang2020,Yoda2020}.

Vertical mirror symmetry has generally been regarded as a key condition for the formation of accidental BICs in planar photonic systems. Accordingly, the merging and band transition of accidental BICs have been investigated in vertically symmetric lattices~\cite{JJin2019,Kang2021,Kang2022,SGLee2024RiP}. In practical implementations, however, vertical mirror symmetry is often broken in substrate-supported structures, where the upper side is exposed to air while the lower side is bounded by a dielectric substrate~\cite{Glowadzka2021,Schiattarella2024}. In such vertically asymmetric lattices, accidental BICs are generally transformed into leaky resonances with finite radiative $Q$ factors. However, the mechanism by which vertical asymmetry prevents complete cancellation of radiative coupling and thereby limits the attainable radiative $Q$ factor remains unclear. Interestingly, it has recently been reported that accidental BICs can be recovered in vertically asymmetric lattices by introducing an additional dielectric layer~\cite{Bai2025}, although a clear physical explanation for this recovery has yet to be established. Very recently, Huang \emph{et al.} showed within a two-mode Friedrich--Wintgen framework that off-$\Gamma$ BICs in substrate-supported photonic-crystal slabs can be recovered by a second odd-parity perturbation that cancels the substrate-induced one~\cite{Huang2026}. The recovery of accidental BICs in vertically asymmetric lattices enables the realization of ultrahigh-$Q$ resonances in substrate-supported photonic structures without requiring free-standing membranes or symmetric embedding, thereby offering greater flexibility in material integration and device design. More fundamentally, such BIC recovery indicates that vertical mirror symmetry is not itself essential for complete radiation cancellation.

In this study, we elucidate the physical mechanism underlying the disappearance and recovery of accidental BICs in vertically asymmetric photonic lattices. A previous semianalytical coupled-mode analysis has shown that accidental BICs arise when the radiation overlap integral of a leaky Bloch mode vanishes~\cite{SGLee2023,SGLee2024JAP}. Here, we further develop this analysis and show that in a vertically symmetric structure, the transverse radiation-source profile has definite even or odd parity, so that the radiation overlap contains only the corresponding parity contribution and can vanish at a particular lattice thickness, producing an accidental BIC. In contrast, in a vertically asymmetric lattice, the radiation-source profile contains both even- and odd-parity components. The two parity contributions enter the complex radiation overlap in quadrature and therefore cannot cancel each other. A true accidental BIC thus requires both contributions to vanish simultaneously. We show that the waveguide dispersion relation prevents their simultaneous vanishing, thereby precluding complete radiation cancellation in a vertically asymmetric lattice. Our analysis reveals that the essential condition for an accidental BIC is not geometric vertical mirror symmetry, but definite parity of the transverse radiation-source profile. By introducing a homogeneous parity-restoring layer (PRL), definite parity of the radiation-source profile can be recovered while the entire structure remains geometrically asymmetric. Under this condition, the remaining parity contribution can be tuned to zero, thereby restoring a true accidental BIC. Finite-element method (FEM) simulations confirm the analytically predicted BIC recovery for both $\mathrm{TE}_0$ and $\mathrm{TE}_1$ Bloch modes in weakly modulated lattices. Additional FEM simulations demonstrate that the same BIC-recovery mechanism remains effective in a strongly modulated Si/SiN binary dielectric grating (BDG) composed of practical dielectric materials. Our results establish radiation-source parity as the essential condition for accidental BIC formation in slab-type photonic lattices and provide fundamental physical insight into BIC formation beyond geometric vertical mirror symmetry.

\section{Lattice structures and perspective}
\begin{figure*}[!t]
\centering \includegraphics[width=\textwidth]{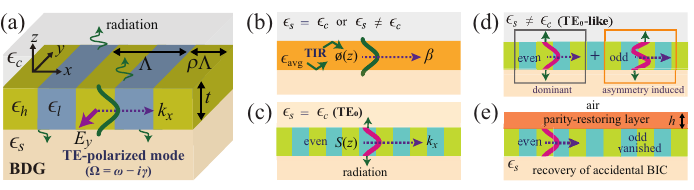}
\caption{(a) Schematic of the 1D binary dielectric grating considered in this study. (b) Guided mode of the unmodulated reference slab, characterized by the transverse mode profile $\phi(z)$ and propagation constant $\beta$. (c) Vertically symmetric lattice, in which the $\mathrm{TE}_0$ radiation-source profile has definite even parity. (d) Vertically asymmetric lattice, where the radiation-source profile consists of a dominant even-parity component and an asymmetry-induced odd-parity component. (e) Vertically asymmetric lattice with a homogeneous parity-restoring layer that restores definite even parity to the radiation-source profile and enables accidental BIC recovery.}
\label{fig1} 
\end{figure*}

Figure~\ref{fig1}(a) shows the representative 1D BDG considered in this study. The BDG consists of periodically arranged high- and low-dielectric-constant regions with dielectric constants $\epsilon_h$ and $\epsilon_l$. The grating has a period $\Lambda$ and a thickness $t$, and the width of the high-dielectric-constant region is $\rho\Lambda$, where $\rho$ is the duty cycle. The lattice is periodic along the $x$ direction and invariant along the $y$ direction, and is bounded by a cover and a substrate with dielectric constants $\epsilon_c$ and $\epsilon_s$, respectively. The structure can support multiple $\mathrm{TE}_n$ and $\mathrm{TM}_n$ guided modes, and each guided mode has its own photonic band gaps at the Bragg condition $\beta=qK/2$~\cite{Ding2007,SGLee2019OL}, where $\beta$ is the propagation constant of the guided mode, $K=2\pi/\Lambda$ is the grating vector, and $q$ is the Bragg order. In this study, we primarily investigate the radiative properties of TE$_0$ Bloch modes near the second band gap using coupled-mode analysis to elucidate the mechanism of out-of-plane radiation in the BDG and the condition for accidental BIC formation, and subsequently demonstrate that the same mechanism applies to TE$_1$ Bloch modes.

Figure~\ref{fig1}(b) shows the unmodulated slab used as the reference waveguide in the coupled-mode analysis. Regardless of whether the slab is vertically symmetric or asymmetric, it supports lossless guided modes confined by total internal reflection (TIR), characterized by the transverse mode profile $\phi(z)$ and propagation constant $\beta$. In our coupled-mode analysis, the dielectric constant of the slab is given by the spatially averaged dielectric constant of the grating layer, $\epsilon_{\mathrm{avg}}=\epsilon_l+\rho\Delta\epsilon$, where $\Delta\epsilon=\epsilon_h-\epsilon_l$. In the periodic lattice shown in Figs.~\ref{fig1}(c) and \ref{fig1}(d), the dielectric modulation scatters these guided modes into the out-of-plane radiation continuum. In this scattering process, the radiation source, whose transverse profile is proportional to the transverse mode profile, $S(z)\propto\phi(z)$, exists only within the grating layer and generates the out-of-plane radiative waves. In the vertically symmetric lattice with $\epsilon_s=\epsilon_c$, the radiation-source profile $S(z)$ has definite even parity for the $\mathrm{TE}_0$ mode, as shown in Fig.~\ref{fig1}(c). When the upper and lower claddings have different dielectric constants ($\epsilon_s\neq\epsilon_c$), the source profile consists of a dominant even-parity component and an asymmetry-induced odd-parity component, as illustrated in Fig.~\ref{fig1}(d). It is worth noting that, for a typical air--SiO$_2$ cladding configuration, the asymmetry-induced odd-parity component is much smaller than the original even-parity component.

It is well known that accidental BICs can be formed in vertically symmetric lattices by tuning the lattice thickness. For the $\mathrm{TE}_0$ mode in the vertically symmetric lattice, coupled-mode analysis shows that an accidental BIC occurs when the radiation overlap associated with the even-parity radiation source vanishes. In a vertically asymmetric lattice, the asymmetry-induced odd-parity component gives rise to residual out-of-plane radiation even when the dominant even-parity contribution is suppressed, thereby limiting the attainable radiative $Q$ factor. Based on the coupled-mode analysis, we demonstrate that an accidental BIC can be recovered even in a vertically asymmetric lattice by restoring the parity of the radiation-source profile. Figure~\ref{fig1}(e) illustrates this concept using a homogeneous PRL. Although the overall lattice remains geometrically asymmetric in the vertical direction, the PRL restores definite even parity to the radiation-source profile within the grating layer. The remaining even-parity radiation overlap can then be tuned to zero, enabling complete radiation cancellation and the recovery of an accidental BIC. Figure~\ref{fig1} illustrates the TE$_0$ case. For the TE$_1$ mode, the roles of the even- and odd-parity components are interchanged.

The results of the coupled-mode analysis, including the predicted recovery of accidental BICs, are verified by FEM simulations. The numerical calculations are performed using the eigenfrequency solver in COMSOL Multiphysics. Because the structures are invariant along the $y$ direction, the calculations are carried out in the 2D $xz$ plane using a computational unit cell of size $\Lambda\times13\Lambda$. Floquet periodic boundary conditions are imposed along the $x$ direction, and perfectly matched layers are used along the $z$ direction to absorb outgoing radiation. A physics-controlled mesh with the element size set to Extremely fine is employed. The complex eigenfrequencies $\Omega=\omega-i\gamma$ of the TE-polarized Bloch modes are obtained from the eigenfrequency calculations, and the corresponding radiative quality factors are evaluated as $Q=\omega/(2\gamma)$.

\section{Results and discussion}
\subsection{Coupled-mode description of out-of-plane radiation}

To describe the out-of-plane radiation of TE-polarized Bloch modes in 1D photonic lattices, we start from the following scalar wave equation~\cite{Yariv1984}
\begin{equation}
\left(\frac{\partial^2}{\partial x^2}+\frac{\partial^2}{\partial z^2}\right)E_y(x,z)+\epsilon(x,z)k_0^2E_y(x,z)=0,
\label{eq:wave-equation}
\end{equation}
where $k_0$ denotes the free-space wavenumber. For photonic lattices satisfying $\epsilon(x,z)=\epsilon(-x,z)$, the periodic dielectric function can be expanded as an even cosine Fourier series
\begin{equation}
\epsilon(x,z)=\epsilon_0(z)+\sum_{n=1}^{\infty}\epsilon_n(z)\cos(nKx),
\label{eq:dielectric-fourier-expansion}
\end{equation}
where within the grating layer, the Fourier coefficients are given by $\epsilon_0=\epsilon_{\mathrm{avg}}$ and $\epsilon_n=(2\Delta\epsilon/n\pi)\sin(n\pi\rho)$. The Fourier coefficients $\epsilon_n(z)$ for $n\geq1$ vanish outside the grating layer. The electric field $E_y(x,z)$ is expressed in Bloch form as
\begin{equation}
E_y(x,z)=\sum_m E_m(z)e^{i(k_x+mK)x},
\label{eq:bloch-field-expansion}
\end{equation}
where $k_x$ is the Bloch wavevector and $E_m(z)$ denotes the transverse field profile of the $m$th spatial harmonic.

Near the second stop band, following the semianalytical Kazarinov–Henry model~\cite{Kazarinov1985}, we retain only the zeroth, first, and second Fourier components of the dielectric function, $\epsilon_0$, $\epsilon_1$, and $\epsilon_2$. The dominant $m=\pm1$ guided harmonics are approximated as $E_{+1}(z)\simeq A\phi(z)$ and $E_{-1}(z)\simeq B\phi(z)$, where $\phi(z)$ is the transverse guided-mode profile of the unmodulated reference slab shown in Fig.~\ref{fig1}(b), whose dielectric constant is given by $\epsilon_0=\epsilon_{\mathrm{avg}}$. The electric field is then approximated as
\begin{equation}
E_y(x,z)=\left(Ae^{iKx}+Be^{-iKx}\right)\phi(z)+E_{\mathrm{rad}}(x,z),
\label{eq:approximated-field}
\end{equation}
where $E_{\mathrm{rad}}(x,z)$ represents the diffracted field induced by the periodic modulation. Substituting this approximation and the truncated dielectric function into Eq.~(\ref{eq:wave-equation}), the zeroth-order radiative component satisfies
\begin{equation}
\left[\frac{d^2}{dz^2}+\epsilon_0(z)k_0^2-k_x^2\right]E_{\mathrm{rad}}(z)=S(z),
\label{eq:radiative-wave-equation}
\end{equation}
where the effective radiation source is given by
\begin{equation}
S(z)=-\frac{k_0^2}{2}\epsilon_1(z)(A+B)\phi(z).
\label{eq:radiation-source}
\end{equation}
Because $\epsilon_1(z)$ is nonzero only within the grating layer, the radiation source is confined to this region and its transverse profile is proportional to $\phi(z)$. The radiative field generated by the source $S(z)$ is expressed using the outgoing Green function $G(z,z')$ of the unmodulated reference slab~\cite{Tomas1995} as
\begin{equation}
E_{\mathrm{rad}}(z)=\int_{-t/2}^{t/2}G(z,z')S(z')\,dz'.
\label{eq:radiative-field}
\end{equation}
Since the scattering by the dielectric modulation occurs only within the grating layer, the radiation field is determined by the source distribution over $-t/2<z<t/2$, where $S(z)$ differs from $\phi(z)$ only by the $z$-independent factor $-(k_0^2/2)\epsilon_1(A+B)$. We therefore refer to $\phi(z)$ within the modulated layer as the reduced transverse radiation-source profile and denote it by $u(z)$, which is given by
\begin{equation}
u(z)=F\cos(\alpha z+\psi), \qquad -t/2<z<t/2,
\label{eq:reduced-source-profile}
\end{equation}
where $\alpha$ is the transverse wavenumber, $\psi$ is the phase shift, and $F$ is the normalization factor satisfying $\int_{-\infty}^{\infty}|\phi(z)|^2\,dz=1$.

\begin{figure*}[!t]
\centering \includegraphics[width=\textwidth]{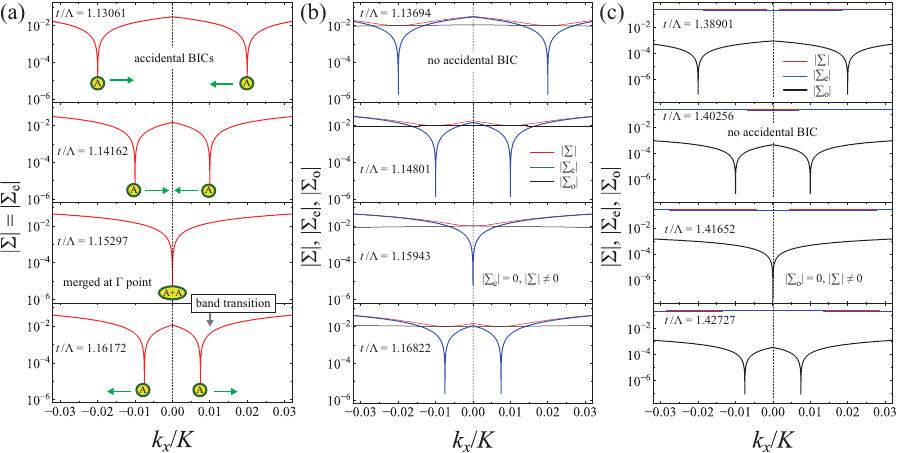}
\caption{Radiation-overlap characteristics of the $\mathrm{TE}_0$ Bloch modes for varying lattice thickness. (a) Vertically symmetric lattice, where $\Sigma_{\mathrm{o}}=0$ and $|\Sigma|=|\Sigma_{\mathrm{e}}|$. The accidental-BIC pair moves toward the $\Gamma$ point, merges at $k_x=0$, and subsequently undergoes a band transition. (b) Vertically asymmetric lattice showing $|\Sigma|$, $|\Sigma_{\mathrm{e}}|$, and $|\Sigma_{\mathrm{o}}|$. The zeros of $\Sigma_{\mathrm{e}}$ merge at the $\Gamma$ point while $\Sigma_{\mathrm{o}}$ remains finite. (c) Evolution of the zeros of $\Sigma_{\mathrm{o}}$, which merge at a different lattice thickness while $\Sigma_{\mathrm{e}}$ remains finite. Because $|\Sigma|\approx|\Sigma_{\mathrm{e}}|\gg|\Sigma_{\mathrm{o}}|$, the curves of $|\Sigma|$ and $|\Sigma_{\mathrm{e}}|$ nearly overlap. For each panel, the four slab thicknesses are selected such that the relevant zero, $|\Sigma|$ in (a), $|\Sigma_{\mathrm e}|$ in (b), and $|\Sigma_{\mathrm o}|$ in (c), occurs at $k_x/K=\pm0.02$, $\pm0.01$, $0$, and $\pm0.007$. The calculations are performed for $\epsilon_{\mathrm{avg}}=8$, $\Delta\epsilon=2$, and $\rho=0.4$. For the vertically symmetric lattice, $\epsilon_c=\epsilon_s=2.1316$. For the vertically asymmetric lattice, $\epsilon_c=1$ and $\epsilon_s=2.1316$.}
\label{fig2} 
\end{figure*}

In the homogeneous cover and substrate, the radiative field takes the form of outgoing plane waves,
\begin{equation}
E_{\mathrm{rad}}(z)=
\begin{cases}
a^{\uparrow}e^{i\mu_c(z-t/2)}, & z>t/2,\\
a^{\downarrow}e^{-i\mu_s(z+t/2)}, & z<-t/2,
\end{cases}
\label{eq:outgoing-radiation}
\end{equation}
where $\mu_c=\sqrt{\epsilon_c k_0^2-k_x^2}$ and $\mu_s=\sqrt{\epsilon_s k_0^2-k_x^2}$ are the out-of-plane wavenumbers in the cover and substrate, respectively, and $a^{\uparrow}$ and $a^{\downarrow}$ denote the corresponding upward and downward radiation amplitudes. The upward and downward radiation amplitudes obtained from Eq.~(\ref{eq:radiative-field}) are given by
\begin{align}
a^{\uparrow}={}&-\frac{k_0^2\tau_c e^{i\mu t/2}}{2D}(A+B)\epsilon_1 \notag\\
&\quad\times\int_{-t/2}^{t/2}u(z)
\left[e^{-i\mu z}+r_s e^{i\mu t}e^{i\mu z}\right]dz,
\label{eq:radiation-amplitude-up}\\
a^{\downarrow}={}&-\frac{k_0^2\tau_s e^{i\mu t/2}}{2D}(A+B)\epsilon_1 \notag\\
&\quad\times\int_{-t/2}^{t/2}u(z)
\left[e^{i\mu z}+r_c e^{i\mu t}e^{-i\mu z}\right]dz,
\label{eq:radiation-amplitude-down}
\end{align}
where $\mu=\sqrt{\epsilon_0k_0^2-k_x^2}$ is the out-of-plane wavenumber in the grating layer, $r_c=(\mu-\mu_c)/(\mu+\mu_c)$ and $r_s=(\mu-\mu_s)/(\mu+\mu_s)$ are the reflection coefficients at the grating-cover and grating-substrate interfaces, respectively, $\tau_c=2\mu/(\mu+\mu_c)$ and $\tau_s=2\mu/(\mu+\mu_s)$ are the corresponding transmission coefficients, and $D=2i\mu(1-r_cr_se^{2i\mu t})$.
To express the radiation amplitudes in terms of the radiation overlap, we define
\begin{equation}
\Sigma=\int_{-t/2}^{t/2}e^{i\mu z}u(z)\,dz.
\label{eq:radiation-overlap}
\end{equation}
Because $u(z)$ is real, the complex conjugate of the radiation overlap is given by $\Sigma^*=\int_{-t/2}^{t/2}u(z)e^{-i\mu z}\,dz$. The radiation amplitudes can therefore be expressed as
\begin{align}
a^{\uparrow}={}&-\frac{k_0^2\tau_c e^{i\mu t/2}}{2D}(A+B)\epsilon_1
\left[\Sigma^*+r_s e^{i\mu t}\Sigma\right],
\label{eq:radiation-amplitude-up-sigma}\\
a^{\downarrow}={}&-\frac{k_0^2\tau_s e^{i\mu t/2}}{2D}(A+B)\epsilon_1
\left[\Sigma+r_c e^{i\mu t}\Sigma^*\right].
\label{eq:radiation-amplitude-down-sigma}
\end{align}
Thus, when $\Sigma=0$, both the upward and downward radiation amplitudes vanish simultaneously. According to Eqs.~(\ref{eq:radiation-amplitude-up-sigma}) and (\ref{eq:radiation-amplitude-down-sigma}), when $A+B=0$, both radiation amplitudes vanish independently of $\Sigma$, yielding a symmetry-protected BIC at one of the two band edges~\cite{SGLee2019PRB}. Away from this symmetry-protected case, $A+B$ is nonzero, and accidental BICs can occur at generic Bloch wavevectors when $\Sigma=0$. For a given Bloch wavevector and eigenfrequency, only the diffracted component with the out-of-plane wavevector $\mu=\sqrt{\epsilon_0k_0^2-k_x^2}$ can propagate out of the slab as zeroth-order radiation. The radiation overlap $\Sigma$ is the Fourier projection of the reduced transverse radiation-source profile $u(z)$ onto the radiating spatial frequency associated with this single propagating diffraction order. Its vanishing therefore eliminates both the upward and downward radiation, even when the cover and substrate have different dielectric constants.

\subsection{Parity decomposition of the radiation overlap}

We now show why vertical asymmetry precludes the formation of accidental BICs. To clarify how vertical asymmetry affects the radiation-overlap condition, we decompose the reduced transverse radiation-source profile $u(z)$ into its even- and odd-parity components as
\begin{equation}
u(z)=u_{\mathrm{e}}(z)+u_{\mathrm{o}}(z).
\label{eq:parity-decomposition}
\end{equation}
The even and odd components are given by $u_{\mathrm{e}}(z)=F\cos\psi\cos(\alpha z)$ and $u_{\mathrm{o}}(z)=-F\sin\psi\sin(\alpha z)$, respectively. Substituting Eq.~(\ref{eq:parity-decomposition}) into Eq.~(\ref{eq:radiation-overlap}) and using the parity of the sine and cosine functions, the radiation overlap can be written as
\begin{equation}
\Sigma=\Sigma_{\mathrm{e}}+i\Sigma_{\mathrm{o}},
\label{eq:overlap-parity-decomposition}
\end{equation}
where
\begin{align}
\Sigma_{\mathrm{e}}&=F\cos\psi\,(M+P),
\label{eq:even-overlap}\\
\Sigma_{\mathrm{o}}&=F\sin\psi\,(M-P),
\label{eq:odd-overlap}
\end{align}
with $P=\sin[(\mu-\alpha)t/2]/(\mu-\alpha)$ and $M=\sin[(\mu+\alpha)t/2]/(\mu+\alpha)$.

Both $\Sigma_{\mathrm{e}}$ and $\Sigma_{\mathrm{o}}$ are real, and the condition $\Sigma=0$ therefore requires $\Sigma_{\mathrm{e}}=0$ and $\Sigma_{\mathrm{o}}=0$ simultaneously. For a vertically symmetric lattice, the reduced transverse radiation-source profile has a definite parity. For the even $\mathrm{TE}_0$ mode considered here, $u_{\mathrm{o}}(z)=0$, and consequently $\Sigma_{\mathrm{o}}=0$ identically. The radiation-cancellation condition $\Sigma=0$ therefore reduces to the single condition $\Sigma_{\mathrm{e}}=0$. By varying the slab thickness, $\Sigma_{\mathrm{e}}$ can cross zero, leading to the formation of an accidental BIC. Figure~\ref{fig2}(a) shows the evolution of $|\Sigma|=|\Sigma_{\mathrm{e}}|$ as a function of $k_x$ for several slab thicknesses. At $t/\Lambda=1.13061$, $|\Sigma|$ vanishes at $k_x/K=\pm0.02$, indicating the formation of a pair of accidental BICs on the upper band. As the slab thickness increases, the two BICs move toward each other and merge at the $\Gamma$ point at $t/\Lambda=1.15297$. With a further increase in the slab thickness, the merged BIC undergoes a band transition to the lower band, after which a pair of accidental BICs re-emerges and moves away from the $\Gamma$ point.

In a vertically asymmetric lattice, however, the reduced transverse radiation-source profile contains both even- and odd-parity components. Simultaneous cancellation of $\Sigma_{\mathrm{e}}$ and $\Sigma_{\mathrm{o}}$ therefore requires $M=P=0$. From the definitions of $M$ and $P$, this requires
\begin{equation}
\alpha t=N\pi,
\label{eq:MP-zero-conditions}
\end{equation}
where $N$ is an integer. On the other hand, the guided-mode dispersion relation of the slab waveguide is~\cite{Yariv1984}
\begin{equation}
\alpha t=\ell\pi+\tan^{-1}\left(\frac{p_c}{\alpha}\right)+\tan^{-1}\left(\frac{p_s}{\alpha}\right),
\label{eq:asymmetric-slab-dispersion}
\end{equation}
\begin{figure*}[!t]
\centering \includegraphics[width=0.85\textwidth]{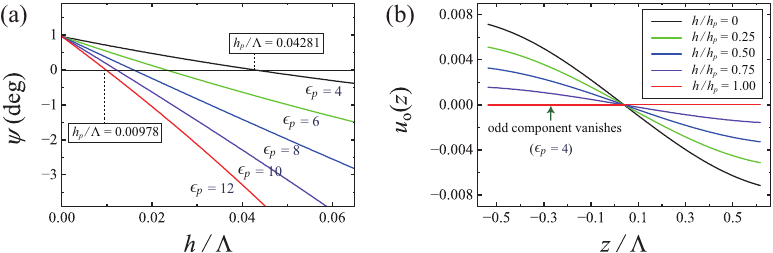}
\caption{Restoration of the reduced radiation-source parity by the PRL. (a) Phase shift $\psi$ as a function of the normalized PRL thickness $h/\Lambda$ for several values of $\epsilon_p$. The intersections with $\psi=0$ determine the corresponding parity-restoring thicknesses $h_p$. (b) Odd-parity component $u_{\mathrm{o}}(z)$ within the grating layer for $\epsilon_p=4$ at selected values of $h$. As $h$ approaches $h_p=0.04281\Lambda$, the odd-parity component gradually diminishes and vanishes throughout the grating layer at $h=h_p$. The calculations are performed at the $\Gamma$ point using $t/\Lambda=1.15297$, the slab thickness of the merged accidental BIC in the vertically symmetric lattice of Fig.~\ref{fig2}(a). For the vertically asymmetric PRL-loaded structure considered here, $\epsilon_{\mathrm{avg}}=8$, $\epsilon_c=1$, and $\epsilon_s=2.1316$.}
\label{fig3} 
\end{figure*}

where $p_c=\sqrt{\beta^2-\epsilon_c k_0^2}$ and $p_s=\sqrt{\beta^2-\epsilon_s k_0^2}$ are the evanescent decay constants in the cover and substrate, respectively, and $\ell=0,1,2,\ldots$ denotes the transverse mode order. The terms $\tan^{-1}(p_c/\alpha)$ and $\tan^{-1}(p_s/\alpha)$ represent the phase contributions associated with TIR at the upper and lower interfaces of the slab, respectively. For the $\mathrm{TE}_0$ mode considered here, $\ell=0$. The same boundary conditions also determine the asymmetry-induced phase shift as
\begin{equation}
\psi=\frac{1}{2}\left[\tan^{-1}\left(\frac{p_c}{\alpha}\right)-\tan^{-1}\left(\frac{p_s}{\alpha}\right)\right],
\label{eq:psi}
\end{equation}
which gives $\psi=0$ for $\epsilon_c=\epsilon_s$, and hence $u_{\mathrm{o}}(z)=0$ in the vertically symmetric case. Since $0<\tan^{-1}(p_c/\alpha)<\pi/2$ and $0<\tan^{-1}(p_s/\alpha)<\pi/2$, the dispersion relation requires $\ell\pi<\alpha t<(\ell+1)\pi$. Hence, $\alpha t$ cannot equal $N\pi$, and $M$ and $P$ cannot vanish simultaneously. Consequently, a true accidental BIC cannot be formed, and the mode remains radiative with a finite radiative $Q$ factor. Because this inequality holds for any transverse mode order $\ell$, the same conclusion applies to higher-order TE modes. Figures~\ref{fig2}(b) and \ref{fig2}(c) illustrate the thickness-dependent evolution of $|\Sigma|$, $|\Sigma_{\mathrm e}|$, and $|\Sigma_{\mathrm o}|$ in the vertically asymmetric lattice. In Fig.~\ref{fig2}(b), for appropriate slab thicknesses, $|\Sigma_{\mathrm e}|$ vanishes at certain wave vectors, indicating complete cancellation of the even-parity radiation contribution. However, $|\Sigma_{\mathrm o}|$ remains finite at the same wave vectors, and therefore the total radiation overlap $|\Sigma|$ does not vanish. Similarly, in Fig.~\ref{fig2}(c), $|\Sigma_{\mathrm o}|$ vanishes at different slab thicknesses, while $|\Sigma_{\mathrm e}|$ remains finite, and $|\Sigma|$ therefore remains nonzero. The curves of $|\Sigma|$ in Fig.~\ref{fig2}(a) and $|\Sigma_{\mathrm e}|$ in Fig.~\ref{fig2}(b) are nearly identical because the quantities that determine them, $\alpha$, $\mu$, and $t$, are nearly the same in the two lattices. Replacing $\epsilon_c=2.1316$ with $\epsilon_c=1$ changes $\alpha$ of the strongly confined TE$_0$ mode by about 1\%, whereas $\mu$ is nearly unchanged. The slab thicknesses in Fig.~\ref{fig2}(b) are about 0.6\% larger than those in Fig.~\ref{fig2}(a), and the phase shift is small, $\psi=0.017$~rad, so that $\cos\psi\simeq1$. Thus, although either parity contribution can be individually canceled by varying the slab thickness, the two contributions cannot vanish simultaneously.

\subsection{Restoration of radiation-source parity}

The parity decomposition of the radiation overlap indicates that the vanishing of $\Sigma$ is not directly tied to the geometric vertical symmetry of the lattice. In a vertically asymmetric lattice, the asymmetry-induced odd-parity component gives rise to a finite $\Sigma_{\mathrm{o}}$, while the even-parity contribution $\Sigma_{\mathrm{e}}$ can still be tuned to zero by varying the slab thickness, as shown in Fig.~\ref{fig2}(b). Therefore, if the reduced transverse radiation-source profile $u(z)$ is restored to even parity within the grating layer, $\Sigma_{\mathrm{o}}$ vanishes identically, and $\Sigma$ can be brought to zero by tuning $\Sigma_{\mathrm{e}}$. To realize this condition, we introduce a homogeneous PRL of thickness $h$ and dielectric constant $\epsilon_p$ above the grating layer, as illustrated in Fig.~\ref{fig1}(e).

To determine how the PRL modifies the parity of the reduced radiation-source profile, we consider the corresponding four-layer reference waveguide consisting of the cover, PRL, guiding layer, and substrate. The presence of the PRL does not change the functional form of the reduced transverse radiation-source profile within the guiding layer, which remains described by $u(z)=F\cos(\alpha z+\psi)$ as in Eq.~(\ref{eq:reduced-source-profile}). The PRL, however, modifies the phase shift $\psi$ through the altered boundary condition at the upper interface of the guiding layer. The modified phase shift is given by
\begin{equation}
\psi=\frac{1}{2}\left[\tan^{-1}\left(\frac{p_{\mathrm{eff}}}{\alpha}\right)-\tan^{-1}\left(\frac{p_s}{\alpha}\right)\right],
\label{eq:psi-prl}
\end{equation}
where $p_{\mathrm{eff}}$ denotes the effective boundary parameter at the upper interface and is given by
\begin{equation}
p_{\mathrm{eff}}=\frac{p_c\cos(\alpha_p h)-\alpha_p\sin(\alpha_p h)}{\cos(\alpha_p h)+(p_c/\alpha_p)\sin(\alpha_p h)},
\label{eq:effective-decay}
\end{equation}
where $\alpha_p=\sqrt{\epsilon_p k_0^2-\beta^2}$ is the transverse wavenumber in the PRL. With the condition $p_{\rm eff}(h_p)=p_s$, the explicit expression for the parity-restoring thickness $h_p$ depends on whether $\alpha_p$ is real or imaginary. For real $\alpha_p$, $h_p$ is given by
\begin{equation}
h_p=\frac{1}{\alpha_p}\left[\tan^{-1}\left(\frac{\alpha_p(p_c-p_s)}{\alpha_p^2+p_cp_s}\right)+m\pi\right],
\label{eq:hp-real}
\end{equation}
where $m=0,1,2,\ldots$ denotes the solution order. The periodicity of the tangent function permits multiple parity-restoring PRL thicknesses. For imaginary $\alpha_p$, $h_p$ is given by
\begin{equation}
h_p=\frac{i}{\alpha_p}\tanh^{-1}\left[-\frac{i\alpha_p(p_c-p_s)}{\alpha_p^2+p_cp_s}\right].
\label{eq:hp-imag}
\end{equation}
Unlike the case of real $\alpha_p$, the imaginary-$\alpha_p$ case admits a unique positive parity-restoring thickness because the hyperbolic tangent is monotonic. For the vertically asymmetric configuration considered here, where $p_c>p_s$, a finite positive parity-restoring thickness exists for $\epsilon_p>\epsilon_s$, regardless of whether $\alpha_p$ is real or imaginary. As $\epsilon_p$ approaches $\epsilon_s$ from above, $h_p$ increases without bound. Throughout this study, when multiple solutions exist, $h_p$ denotes the smallest positive parity-restoring thickness. Notably, the condition $\epsilon_p>\epsilon_s$ places no upper bound on $\epsilon_p$, so parity restoration remains possible even when the PRL dielectric constant exceeds the spatially averaged dielectric constant of the guiding layer, $\epsilon_p>\epsilon_{\rm avg}$.

\begin{figure*}[!t]
\centering \includegraphics[width=\textwidth]{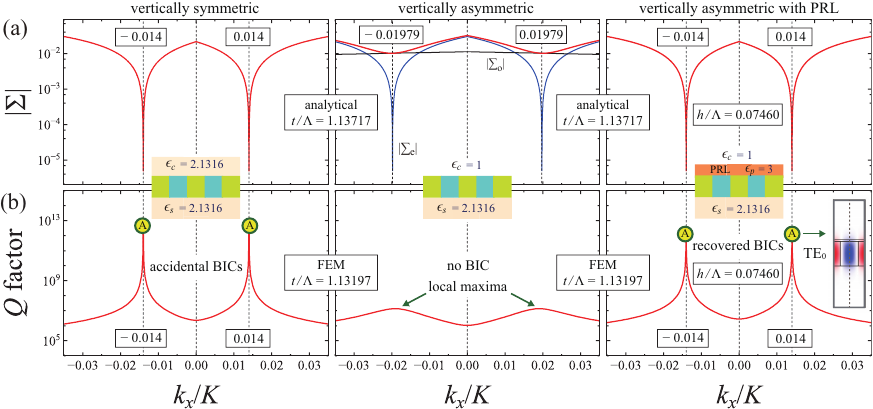}
\caption{Recovery of accidental BICs for the TE$_0$ Bloch mode. (a) Analytical radiation overlap $|\Sigma|$ and (b) FEM-simulated radiative $Q$ factor for the vertically symmetric lattice (left), vertically asymmetric lattice without the PRL (middle), and vertically asymmetric lattice with the PRL (right). In the vertically symmetric lattice, accidental BICs occur at $k_x/K=\pm0.014$. Vertical asymmetry removes the zeros of $|\Sigma|$ and leaves finite $Q$ factors. With the PRL of $\epsilon_p=3$ and $h/\Lambda=0.07460$, the zeros of $|\Sigma|$ and the corresponding high-$Q$ states reappear at the same wave vectors, demonstrating the recovery of the accidental BICs. The inset in the right panel shows the electric-field distribution of the recovered TE$_0$ accidental BIC at $k_x/K=0.014$. The slab thicknesses are $t/\Lambda=1.13717$ for the analytical calculations and $t/\Lambda=1.13197$ for the FEM simulations. The calculations are performed for $\epsilon_{\rm avg}=8$, $\Delta\epsilon=1$, and $\rho=0.55$. For the vertically symmetric lattice, $\epsilon_c=\epsilon_s=2.1316$. For the vertically asymmetric lattices, $\epsilon_c=1$ and $\epsilon_s=2.1316$.}
\label{fig4} 
\end{figure*}

\begin{figure*}[!t]
\centering \includegraphics[width=\textwidth]{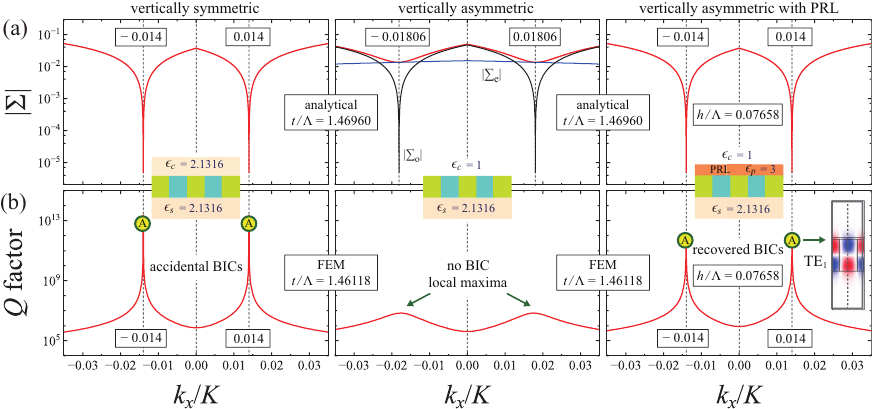}
\caption{Recovery of accidental BICs for the TE$_1$ Bloch mode. (a) Analytical radiation overlap $|\Sigma|$ and (b) FEM-simulated radiative $Q$ factor for the vertically symmetric lattice (left), vertically asymmetric lattice without the PRL (middle), and vertically asymmetric lattice with the PRL (right). In the vertically symmetric lattice, accidental BICs occur at $k_x/K=\pm0.014$. Vertical asymmetry removes the zeros of $|\Sigma|$ and leaves finite $Q$ factors. With the PRL of $\epsilon_p=3$ and $h/\Lambda=0.07658$, the zeros of $|\Sigma|$ and the corresponding high-$Q$ states reappear at the same wave vectors, demonstrating the recovery of the accidental BICs. The inset in the right panel shows the electric-field distribution of the recovered TE$_1$ accidental BIC at $k_x/K=0.014$. The slab thicknesses are $t/\Lambda=1.46960$ for the analytical calculations and $t/\Lambda=1.46118$ for the FEM simulations. The other structural parameters are the same as those in Fig.~\ref{fig4}.}
\label{fig5} 
\end{figure*}

At $h=h_p$, the condition $p_{\rm eff}=p_s$ gives $\psi=0$ through Eq.~(\ref{eq:psi-prl}), and the reduced transverse radiation-source profile becomes $u(z)=F\cos(\alpha z)$, which has definite even parity. Figure~\ref{fig3}(a) shows the variation of $\psi$ with the PRL thickness $h$ for several values of $\epsilon_p$. For each $\epsilon_p$, the first zero of $\psi$ corresponds to the smallest positive parity-restoring thickness $h_p$. The required $h_p$ decreases with increasing $\epsilon_p$, from $h_p/\Lambda=0.04281$ for $\epsilon_p=4$ to $h_p/\Lambda=0.00978$ for $\epsilon_p=12$. Parity restoration is also obtained for $\epsilon_p=10$ and $12$, both of which exceed $\epsilon_{\rm avg}=8$. Figure~\ref{fig3}(b) shows the evolution of the odd-parity component $u_{\mathrm o}(z)$ for $\epsilon_p=4$ as $h$ approaches $h_p$. The magnitude of $u_{\mathrm o}(z)$ progressively decreases with increasing $h$ and vanishes throughout the grating layer at $h=h_p$. This confirms that the PRL restores even parity of the reduced transverse radiation-source profile by eliminating the asymmetry-induced odd-parity source component.

\subsection{Recovery of accidental BICs}
\label{sec:recovery}

We now demonstrate the recovery of accidental BICs in the vertically asymmetric lattice with the PRL on top. Figure~\ref{fig4} compares the analytical and FEM results for the vertically symmetric lattice, the vertically asymmetric lattice without the PRL, and the vertically asymmetric lattice with the PRL. For the vertically symmetric lattice, the slab thickness is set to $t/\Lambda=1.13717$ in the analytical calculation, at which $|\Sigma|$ vanishes at $k_x/K=\pm0.014$, indicating the formation of accidental BICs. In the corresponding FEM simulation, accidental BICs at the same wave vectors are obtained at a slightly different slab thickness of $t/\Lambda=1.13197$. The slight difference in the slab thickness required to realize the accidental BICs at the given wave vectors reflects the approximations of the analytical model, which uses the transverse mode profile of the spatially averaged unmodulated slab as the radiation source, whereas the FEM simulation directly treats the actual periodic lattice. When vertical asymmetry is introduced by replacing the cover with air while keeping the same slab thicknesses, $|\Sigma_{\mathrm e}|$ vanishes at $k_x/K=\pm0.01979$, whereas $|\Sigma_{\mathrm o}|$ remains finite and the total radiation overlap $|\Sigma|$ does not vanish. The FEM-simulated radiative $Q$ factors accordingly remain finite and exhibit only local maxima.

The PRL is then introduced without changing these slab thicknesses, which remain $t/\Lambda=1.13717$ and $t/\Lambda=1.13197$ for the analytical calculation and FEM simulation, respectively. For $\epsilon_p=3$, the parity-restoring PRL thickness of $h/\Lambda=0.07460$ is obtained from the analytical model and used in both the analytical calculation and the FEM simulation. At this PRL thickness, Fig.~\ref{fig4}(a) shows that the zeros of $|\Sigma|$ reappear at $k_x/K=\pm0.014$ in the analytical calculation. The FEM simulation shown in Fig.~\ref{fig4}(b) confirms the analytically predicted recovery of the accidental BICs, yielding radiative $Q$ factors exceeding $10^{11}$ at the same wave vectors. The inset in Fig.~\ref{fig4}(b) shows the electric-field distribution of the recovered TE$_0$ accidental BIC at $k_x/K=0.014$. At the parity-restoring condition $\psi=0$, the PRL yields a transverse radiation-source profile $u'_{\mathrm e}(z)=F'\cos(\alpha z)$, which is proportional to the source profile $u_{\mathrm e}(z)=F\cos(\alpha z)$ of the vertically symmetric three-layer structure. Since the multiplicative factors $F'$ and $F$ do not affect the zero condition of the radiation overlap, the out-of-plane radiation vanishes at the same slab thickness $t$ for both the vertically symmetric lattice and the vertically asymmetric lattice with the PRL.

We next examine the recovery of accidental BICs for the TE$_1$ Bloch mode. In the three-layer structure without the PRL, the reduced transverse radiation-source profile of the TE$_1$ mode satisfies the following phase-shift relation:
\begin{equation}
\psi=\frac{\pi}{2}+\frac{1}{2}\left[\tan^{-1}\left(\frac{p_c}{\alpha}\right)-\tan^{-1}\left(\frac{p_s}{\alpha}\right)\right].
\end{equation}
The additional $\pi/2$ term originates from the odd transverse mode order, $\ell=1$, in Eq.~(\ref{eq:asymmetric-slab-dispersion}). In the vertically symmetric lattice, $p_c=p_s$, and therefore $\psi=\pi/2$. The reduced transverse radiation-source profile is then given by $u(z)=-F\sin(\alpha z)$, which has definite odd parity. Consequently, $\Sigma_{\mathrm e}=0$ and $|\Sigma|=|\Sigma_{\mathrm o}|$. Figure~\ref{fig5}(a) shows that $|\Sigma|$ vanishes at $k_x/K=\pm0.014$ for $t/\Lambda=1.46960$, producing a pair of accidental BICs. The FEM simulation in Fig.~\ref{fig5}(b) yields accidental BICs at the same wave vectors for $t/\Lambda=1.46118$. When vertical asymmetry is introduced by replacing the cover with air while keeping the same slab thicknesses, $p_c\neq p_s$ and the phase shift deviates from $\pi/2$, introducing an asymmetry-induced even-parity component into the radiation-source profile. In this case, $|\Sigma_{\mathrm o}|$ vanishes at $k_x/K=\pm0.01806$, whereas $|\Sigma_{\mathrm e}|$ remains finite and the total radiation overlap $|\Sigma|$ does not vanish. The FEM-simulated radiative $Q$ factors accordingly remain finite and exhibit only local maxima.

For the vertically asymmetric TE$_1$ lattice, the analytical model gives a parity-restoring PRL thickness of $h/\Lambda=0.07658$ for $\epsilon_p=3$. With the PRL, the upper-interface boundary parameter $p_c$ in the TE$_1$ phase-shift relation is replaced by $p_{\rm eff}$. At this thickness, $p_{\rm eff}=p_s$, and the phase shift returns to $\psi=\pi/2$. The asymmetry-induced even-parity component therefore vanishes, restoring the definite odd-parity radiation-source profile $u(z)=-F\sin(\alpha z)$. Consequently, the zeros of $|\Sigma|$ reappear at $k_x/K=\pm0.014$ in the analytical calculation. The FEM simulation confirms the recovery of the accidental BICs at the same wave vectors using the same slab thickness as in the vertically symmetric lattice. The inset in Fig.~\ref{fig5}(b) shows the electric-field distribution of the recovered TE$_1$ accidental BIC at $k_x/K=0.014$. These results demonstrate that the PRL-based recovery mechanism applies to both even- and odd-parity radiation-source profiles.

\begin{figure}[!t]
\centering \includegraphics[width=0.9\columnwidth]{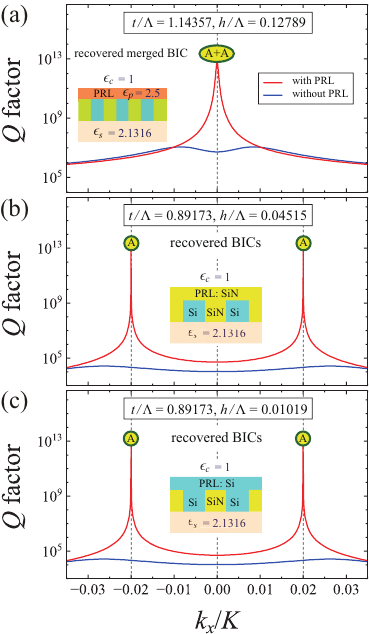}
\caption{Extension of accidental-BIC recovery to a merged BIC and a strongly modulated Si/SiN BDG. (a) FEM-simulated radiative $Q$ factors for the recovery of a merged accidental BIC at the $\Gamma$ point for $t/\Lambda=1.14357$, using a PRL with $\epsilon_p=2.5$ and $h/\Lambda=0.12789$. The other structural parameters in (a) are the same as those in Fig.~\ref{fig4}. (b) Recovery of a pair of accidental BICs at $k_x/K=\pm0.02$ in a strongly modulated Si/SiN BDG with a SiN PRL of $h/\Lambda=0.04515$ at $t/\Lambda=0.89173$. (c) Recovery of the same accidental BICs with a Si PRL of $h/\Lambda=0.01019$ at the same BDG thickness. The blue and red curves represent the structures without and with the PRL, respectively. The Si/SiN BDG in (b) and (c) has a duty cycle of $\rho=0.4$, with $\epsilon_{\rm Si}=12.1104$, $\epsilon_{\rm SiN}=4$, and $\epsilon_{\rm SiO_2}=2.1316$.}
\label{fig6} 
\end{figure}

\begin{table*}[!t]
\centering
\caption{Slab thickness $t$ and PRL thickness $h$ obtained from the analytical model and from FEM simulations at the reference wavevector $k_x$ listed in the table. For each lattice, $t$ is the thickness at which the accidental BIC forms at the reference $k_x$ in the vertically symmetric lattice, and $h$ is the PRL thickness at which the BIC is recovered at the same $k_x$ in the vertically asymmetric lattice. The eigenfrequency of the BIC at the reference $k_x$ is given as $\omega\Lambda/2\pi c$. The values in parentheses are the deviations of the model from FEM. The FEM thicknesses are given to the resolution of the scans ($10^{-5}\Lambda$ or finer).\label{tab1}}
\begin{tabular*}{\textwidth}{@{\extracolsep\fill}lcccccccc@{\extracolsep\fill}}
\toprule
\textbf{Structure} & $k_x/K$ & $\epsilon_p$ & \multicolumn{2}{c}{$t/\Lambda$} & \multicolumn{2}{c}{$\omega\Lambda/2\pi c$} & \multicolumn{2}{c}{$h/\Lambda$} \\
 & & & model & FEM & model & FEM & model & FEM \\
\midrule
TE$_0$, Fig.~\ref{fig4} & 0.014 & 3 & 1.13717 & 1.13197 (0.46\%) & 0.37754 & 0.37811 (0.15\%) & 0.07460 & 0.075170 (0.8\%) \\
TE$_1$, Fig.~\ref{fig5} & 0.014 & 3 & 1.46960 & 1.46118 (0.58\%) & 0.40821 & 0.40922 (0.25\%) & 0.07658 & 0.077400 (1.1\%) \\
SiN PRL, Fig.~\ref{fig6}(b) & 0.02 & 4 & 1.12778 & 0.89173 (26\%) & 0.39897 & 0.43189 (7.6\%) & 0.04263 & 0.045150 (5.6\%) \\
Si PRL, Fig.~\ref{fig6}(c) & 0.02 & 12.1104 & 1.12778 & 0.89173 (26\%) & 0.39897 & 0.43189 (7.6\%) & 0.00964 & 0.010190 (5.4\%) \\
\bottomrule
\end{tabular*}
\end{table*}

Having demonstrated the recovery of off-$\Gamma$ accidental BICs for both the TE$_0$ and TE$_1$ Bloch modes, we further consider the case in which the two BICs merge at the $\Gamma$ point. Merged BICs are particularly important for practical applications because their coalescence enables ultrahigh-$Q$ resonances with enhanced robustness against fabrication imperfections~\cite{JJin2019,Kang2022}. Compared with the isolated accidental BICs shown in Figs.~\ref{fig4}(b) and \ref{fig5}(b), the recovered merged BIC in Fig.~\ref{fig6}(a) exhibits a broader high-$Q$ region around the $\Gamma$ point, indicating enhanced robustness against variations in $k_x$. In Figs.~\ref{fig4}, \ref{fig5}, and \ref{fig6}(a), the recovery of accidental BICs was demonstrated for idealized, weakly modulated BDGs with $\Delta\epsilon=1$. To examine the applicability of the PRL approach beyond the weak-modulation regime, we next consider a strongly modulated BDG composed of Si and SiN. In the vertically symmetric structure with SiO$_2$ as the cover and substrate, we first confirm that the Si/SiN BDG supports a pair of accidental BICs at $k_x/K=\pm0.02$ for $t/\Lambda=0.89173$. When the SiO$_2$ cover is replaced by air, the accidental BICs disappear, as shown by the blue curves in Figs.~\ref{fig6}(b) and \ref{fig6}(c). Upon introducing a SiN PRL with $h/\Lambda=0.04515$, the two accidental BICs are recovered at $k_x/K=\pm0.02$ without changing the BDG thickness, as shown in Fig.~\ref{fig6}(b). Figure~\ref{fig6}(c) shows that the accidental BICs are also recovered with a Si PRL at $h/\Lambda=0.01019$. These results demonstrate that the PRL approach remains effective for a strongly modulated Si/SiN BDG composed of practical dielectric materials and is not restricted to a specific PRL material.

\subsection{Validity, robustness, and scope of parity restoration}
\label{sec:validity}

The two-wave Kazarinov--Henry model is valid for slab-type lattices with weak modulation, $\Delta\epsilon\ll\epsilon_{\mathrm{avg}}$, and near the Bragg condition $\beta\simeq K$, where the leaky Bloch mode is dominated by the $m=\pm1$ harmonics, which are the two counterpropagating guided waves coupled by Bragg diffraction. Table~\ref{tab1} compares the analytical and FEM results at the reference wavevectors of Figs.~\ref{fig4}--\ref{fig6}. At the reference wavevector $k_x/K=0.014$ of Figs.~\ref{fig4} and \ref{fig5}, the slab thicknesses of the accidental BICs obtained from the model and FEM differ by 0.46\% for TE$_0$ and 0.58\% for TE$_1$, whereas the BIC eigenfrequencies differ by 0.15\% and 0.25\%. The PRL thickness given by Eqs.~(\ref{eq:hp-real}) and (\ref{eq:hp-imag}) depends on the permittivities of the cover, substrate, and PRL and on the propagation constant and the frequency of the guided mode. Of these, the permittivities are common to the model and FEM, and the propagation constant and the frequency are nearly the same at the reference wavevector. Consequently, the FEM-optimized PRL thicknesses differ from the analytical values by 0.8\% and 1.1\%. Even at the analytical PRL thicknesses, the FEM $Q$ factors reach $3\times10^{11}$ and $8\times10^{10}$, and they increase to $1.6\times10^{13}$ and $2.5\times10^{12}$ at the FEM-optimized thicknesses. For strongly modulated lattices, such as the Si/SiN grating of Fig.~\ref{fig6}, the model no longer predicts the slab and PRL thicknesses accurately, and these thicknesses in Fig.~\ref{fig6} were determined by FEM simulations. For the recovered BIC of the TE$_0$ mode in Fig.~\ref{fig4}, convergence was verified by scaling the maximum element size by factors of 1.5 to 0.25, corresponding to $6\times10^{4}$ to $1.3\times10^{6}$ degrees of freedom. The PRL thickness at which the radiative $Q$ factor peaks is unchanged, and the $Q$ factors at $h/\Lambda=0.0748$ and $0.0756$, $4\times10^{-4}\Lambda$ away from the peak, agree to within 1\%. The peak values themselves are finite because of the finite numerical accuracy and the finite resolution of the scans, and should be regarded as lower bounds.

\begin{figure*}[!t]
\centering \includegraphics[width=0.85\textwidth]{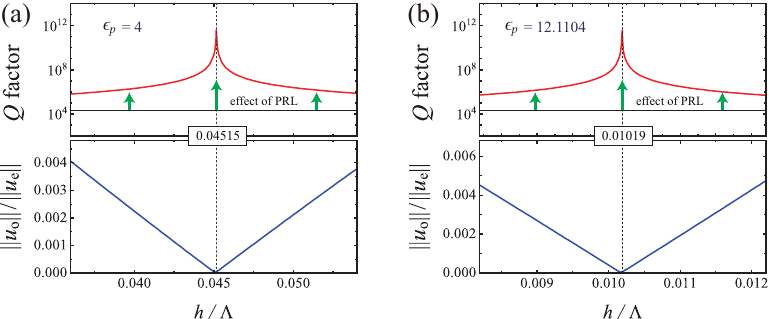}
\caption{Parity restoration in the strongly modulated Si/SiN BDG shown in Fig.~\ref{fig6}. FEM-simulated radiative $Q$ factor (top) and ratio of the odd- to even-parity components of the full radiation-source profile, $\|u_{\mathrm o}\|/\|u_{\mathrm e}\|$ (bottom), as functions of the PRL thickness $h$ for (a) a SiN PRL with $\epsilon_p=4$ and (b) a Si PRL with $\epsilon_p=12.1104$. The vertical dashed lines indicate the FEM-optimized PRL thicknesses, $h/\Lambda=0.04515$ and $0.010190$, respectively. The horizontal lines indicate the radiative $Q$ factor of the vertically asymmetric lattice without the PRL, $Q=2.1\times10^{4}$. The calculations are performed at $k_x/K=0.02$ with $t/\Lambda=0.89173$, and all other parameters are the same as in Fig.~\ref{fig6}.}
\label{fig7}
\end{figure*}

To verify that the recovery in the strongly modulated Si/SiN grating of Fig.~\ref{fig6} is governed by radiation-source parity, we calculated the full radiation-source profile $u(z)\propto\int_0^\Lambda[\epsilon(x,z)-\epsilon_{\mathrm{avg}}]E_y(x,z)e^{-ik_xx}\,dx$ inside the grating layer from the FEM eigenmodes. This profile includes all Bloch harmonics and reduces to $u(z)\propto E_{+1}(z)+E_{-1}(z)$ when only the $m=\pm1$ harmonics are retained. We decomposed $u(z)$ into its even and odd parts, $u_{\mathrm e}(z)$ and $u_{\mathrm o}(z)$, and evaluated the odd-to-even ratio $\|u_{\mathrm o}\|/\|u_{\mathrm e}\|$ with $\|u_{\mathrm{e,o}}\|^2=\int|u_{\mathrm{e,o}}(z)|^2\,dz$ over the grating layer (Fig.~\ref{fig7}). In the vertically symmetric grating, this ratio vanishes, whereas in the air-covered grating it is 2.2\% and limits the radiative $Q$ factor to $2.1\times10^{4}$. With a SiN or Si PRL, the odd-to-even ratio decreases as $h$ approaches the optimal thickness and reaches minima of $3.4\times10^{-5}$ and $4.4\times10^{-5}$ near $h/\Lambda=0.04515$ and $0.010190$, respectively, where the radiative $Q$ factor peaks. These results confirm that the recovery of the accidental BICs in Fig.~\ref{fig6} is governed by radiation-source parity.

The guided mode in the grating layer sees the PRL and the air above it only through the boundary condition at its upper interface, which is characterized by $p_{\rm eff}=-\phi'(t/2)/\phi(t/2)$ in Eq.~(\ref{eq:effective-decay}). Without the PRL, $p_{\rm eff}$ equals the decay constant $p_c$ in air. We therefore define the effective permittivity of the PRL--air region as that of a semi-infinite homogeneous cover in which the guided-mode field decays as $\exp[-p_{\rm eff}(z-t/2)]$, $\epsilon_{c,\mathrm{eff}}=(\beta^2-p_{\rm eff}^2)/k_0^2$. At $h=h_p$, the condition $p_{\rm eff}=p_s$ gives $\epsilon_{c,\mathrm{eff}}=\epsilon_s$ exactly, so that the guided mode sees the same medium above and below the grating layer. This equivalence restores the parity of the radiation-source profile within the grating layer. Outside the grating layer, however, the fields of the guided mode differ above and below, and the mode as a whole remains vertically asymmetric. The accidental BIC is nevertheless recovered because the radiation source exists only within the grating layer.

The ultrahigh $Q$ factors presented here are radiative $Q$ factors of ideal lossless structures. In practice, structural deviations and material absorption limit the achievable $Q$ factor of any BIC. These limitations are not specific to the PRL approach and are expected to be mitigated by advances in fabrication technology and the use of low-loss materials. In contrast, the radiative limit imposed by vertical asymmetry is intrinsic to substrate-supported structures and cannot be overcome by improving fabrication accuracy or material quality. The PRL removes this limit while introducing only one additional structural parameter, its thickness. For the Si/SiN grating with a SiN PRL, the radiative $Q$ factor remains two orders of magnitude higher than the value of $2.1\times10^{4}$ obtained without the PRL for thickness deviations up to $5.1\times10^{-3}\Lambda$, and four orders of magnitude higher for deviations up to $5.1\times10^{-4}\Lambda$. For $\lambda=1550$~nm ($\Lambda=669$~nm), these correspond to tolerances of $\pm3.4$~nm and $\pm0.34$~nm around the optimal thickness of 30.2~nm.

We derived the requirement of definite radiation-source parity for 1D photonic lattices and showed that it holds for different cover and substrate materials and for different transverse mode orders. Moreover, the full-wave parity decomposition in Fig.~\ref{fig7} confirms that the parity requirement remains valid for the strongly modulated Si/SiN grating, beyond the weak-modulation assumption of the analytical model. The derivation is based on the slab geometry, in which the grating layer is bounded by flat upper and lower interfaces and the radiation source can be described by the guided-mode profile of the corresponding unmodulated reference slab. The parity requirement is therefore expected to apply to slab-type photonic lattices in general. Coupled-mode analysis has also been used to study the radiative properties of 2D photonic-crystal slabs~\cite{Yang2014,Liang2011}, which suggests that the present radiation-source-parity analysis can be formally extended to 2D lattices. This extension will be pursued in future work. The PRL approach can be tested experimentally with a Si/SiN grating on a SiO$_2$ substrate and a SiN layer deposited on top. Angle-resolved reflectance or transmission measurements for a series of layer thicknesses spanning the optimal PRL thickness obtained from the FEM simulations would reveal the recovery as a narrowing of the resonance linewidth near this thickness. At a wavelength of 1550~nm, Si, SiN, and SiO$_2$ exhibit low material absorption, while fabrication imperfections may introduce additional scattering and radiation losses that reduce the attainable $Q$ factor. The benefit of the recovered BICs for applications such as lasing, sensing, and nonlinear optics depends on the specific device and is left for future work.

\section{Conclusion}

In conclusion, we have shown that the formation of accidental BICs in planar photonic lattices is governed by the parity of the transverse radiation-source profile rather than by geometric vertical mirror symmetry itself. By extending the semianalytical coupled-mode analysis to vertically asymmetric lattices, we decomposed the radiation overlap into even- and odd-parity contributions. In vertically asymmetric lattices, both contributions are generally present, and the waveguide dispersion prevents them from vanishing simultaneously, thereby precluding a true accidental BIC. We then introduced a homogeneous PRL that eliminates the asymmetry-induced opposite-parity source component and restores definite parity of the reduced transverse radiation-source profile. Under the parity-restoring condition, the remaining parity contribution vanishes at the same slab thickness as in the corresponding vertically symmetric structure, recovering the accidental BIC without retuning the slab thickness. FEM simulations confirmed the predicted recovery of isolated off-$\Gamma$ accidental BICs for both the TE$_0$ and TE$_1$ Bloch modes, as well as a merged accidental BIC at the $\Gamma$ point. The applicability of the PRL approach was further demonstrated in a strongly modulated Si/SiN BDG composed of practical dielectric materials. These results establish radiation-source parity as the essential physical condition for accidental BIC formation in slab-type photonic lattices and provide a route to recovering accidental BICs in vertically asymmetric photonic lattices.

\bigskip
\noindent\textbf{Acknowledgements}\\
This research was supported by a grant from the National Research Foundation of Korea, funded by the Ministry of Science and ICT (No. 2022R1A2C1011091).

\bigskip
\noindent\textbf{Conflict of Interest}\\
The authors declare no conflicts of interest.

\bigskip
\noindent\textbf{Data Availability Statement}\\
Data underlying the results in this paper may be obtained from the authors upon reasonable request.

\bigskip
\noindent\textbf{Keywords}\\
bound states in the continuum, accidental BICs, vertical asymmetry, radiation-source parity, parity-restoring layer, coupled-mode analysis

\end{document}